# Orientation-Controlled Construction of Superstructures of Atomically-Flat Nanocrystals: Pushing the Limits of Ultra-Thin Colloidal Gain Media

Onur Erdem[1], Sina Foroutan[1], Negar Gheshlaghi[1], Burak Guzelturk[2], Yemliha Altintas[1,3], and Hilmi Volkan Demir[1,4*]

[1]Department of Electrical and Electronics Engineering, Department of Physics, UNAM – Institute of Materials Science and Nanotechnology, Bilkent University, Ankara 06800, Turkey

[2]Advanced Photon Source, Argonne National Laboratory, Lemont, IL, USA 60439

[3]Department of Materials Science and Nanotechnology, Abdullah Gul University, Kayseri 38080, Turkey

[4]LUMINOUS! Centre of Excellence for Semiconductor Lighting and Displays, The Photonics Institute, School of Electrical and Electronic Engineering, School of Physical and Mathematical Sciences, School of Materials Science and Engineering, Nanyang Technological University, 50 Nanyang Avenue, Singapore 639798, Singapore
*E-mail: hvdemir@ntu.edu.sg

**We propose and demonstrate a method for the construction of highly uniform, multilayered, orientation-controlled superstructures of CdSe/CdZnS core/shell colloidal nanoplatelets (NPLs) using bi-phase liquid interface. These atomically-flat nanocrystals are sequentially deposited, all face-down onto a solid substrate, into slabs having monolayer-precise thickness and excellent homogeneity over several tens of cm$^2$ areas. Owing to the near-unity surface coverage and film uniformity of this deposition technique, amplified spontaneous emission (ASE) is observed from an uncharacteristically thin colloidal film having only 6 layers of NPLs, which corresponds to a mere 42 nm thickness. Furthermore, systematic studies of optical gain properties of these NPL superstructures constructed having precise numbers of NPL layers tuned from 6 to 15 revealed the reduction in the gain threshold with the increasing number of NPL monolayers, along with a continuous spectral shift in the position of the ASE peak (by ~18 nm). These observations can be well explained by the variation of the optical field confinement factor with the NPL waveguide thickness and propagation wavelength. This work demonstrates the possibility of fabricating thickness-tunable, large-area three-dimensional superstructures made of NPL building blocks, which can be additively constructed one monolayer at a time. The proposed technique can also be extended to build hybrid NPL films of mixed orientations and allow for precise large-area device engineering.**

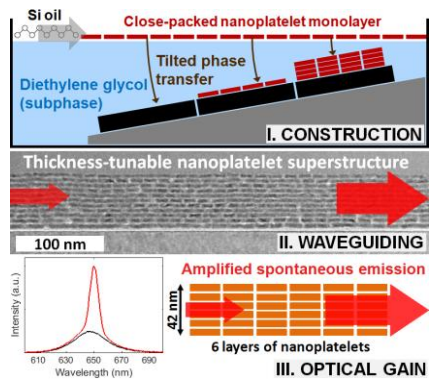

## 1. Introduction

Colloidal semiconductor nanocrystals (NCs) have been thoroughly investigated for the past two decades as optical gain material, largely owing to their favorable properties including wide color tunability, high quantum yield and strong optical abrosption. NCs of different dimensionalities, including quasi-0-dimensional quantum dots, 1-dimensional nanorods (NRs) and 2-dimensional colloidal quantum wells each have thus far been characterized widely in terms of their optical gain performance for their potential use as optical gain media in NC lasers.[1–5] Colloidal quantum wells, also commonly nicknamed as nanoplatelets (NPLs), are the most recent class of these colloidal semiconductor NCs, which offer favorable properties for lasing, including narrow emission linewidth due to their atomically-flat lateral surfaces and enhanced oscillator strength.[6] Soon after their introduction, NPLs have been shown to be an excellent family of nanoemitters for lasing as they feature suppressed inhomogeneous broadening,[6,7] reduced gain thresholds[4,8,9] and escalated gain coefficients.[10] Since then, NPLs have been extensively studied for optical gain and lasing.[4,11–14]

Solid films of nanoemitters typically require a minimum thickness to display optical gain.[15–18] To assure their films have sufficient thickness, they are commonly prepared via spin-coating or drop-casting of a concentrated NC solution onto the substrate. However, drop-casting technique commonly leads to nonuniform films with possibly large variance in thickness across large area films. Although spin-coating generally produces uniform films with a well-defined thickness, controlling the film thickness via experimental parameters such as rotation speed of the spinner and amount and concentration of NCs is challenging in terms of precision and reproducibility, particularly in the case of anisotropic NCs (*i.e.,* NRs and NPLs) that can exhibit distinct in-film orientations.[19,20] As a result, creating thickness-tunable NPL films and studying and understanding their optical gain as a function of the film thickness with high enough precision is cumbersome.

To address these issues, we propose and demonstrate the construction of superstructures of NPLs, all oriented face-down, with a precise number of complete monolayers of these NPLs using a thin film deposition technique which we have developed to create large-scale (several tens of cm$^2$) and thickness-controlled (1- to 15-monolayered) films of NPLs. In this strategy, densely-packed single layers of NPLs are deposited on substrates repetitively via this self-assembly deposition, yielding excellent homogeneity throughout the film. This sequential deposition of NPLs enables controlling the thickness of the resulting films precisely and creating large-area slabs of NPLs with a desired thickness, which can be in principle utilized as optically active structures.

In our previous work,[21] we showed the technique of assembling a single monolayer of NPLs in a distinct orientation of our choice, face-down or edge-up. Here we extend this technique to its successive application with the modification of subphase and the additional configuration of tilted phase transfer to build up layered NPL superstructures, adding one monolayer at a time, until reaching the desired thickness. These are the vertical repetitive structures of self-assembled monolayers of NPLs sitting laterally side by side in close packing. In contrast to the widely studied layer-by-layer deposition technique based on dipping into alternating solutions of positively and negatively charged polymers and/or colloidal nanoparticles (e.g. charged semiconductor NCs),[22,23] linker molecules between NC layers are not required in our approach.

Here, we systematically studied the optical gain properties of these multilayered NPL superstructures while decreasing the number of NPL layers. Thicker NPL films lead to gain easily at reduced gain thresholds. However, as making the film thinner and thinner, it becomes progressively difficult to observe gain. The dense packing and high uniformity of our deposited NPL superstructures on fused silica, on the other hand, enable the observation of optical gain down to 6 layers of such NPL constructs, which is altogether as thin as 42 nm. This film thickness is much smaller than the previously reported thicknesses typically of few 100s nm for the NC films displaying optical gain,[3,10,24,25] and to the best of our knowledge, the thinnest ever reported optical gain media without using any optical feedback on a bare substrate. Moreover, we find that thinner than a critical thickness (depending on material parameters and conditions), it is not possible to induce gain no matter how hard pumped. Therefore, the observation of optical gain is limited to the critical thickness, below which no waveguiding mode within the NPL slab exists. Below this critical thickness, which is thus a cut-off thickness for waveguiding of the NPL slab, the resulting confinement factor is practically zero. Right above the critical thickness, however, there is a sudden major jump of 4 orders of magnitude in the confinement factor. By increasing the number of NPL layers in the film further beyond the critical thickness, we were able to observe gradual decrease in the gain threshold, which can be explained by progressively increasing field confinement factor within the NPL slab.

In this work, equipped with the ability of large-scale bottom-up construction of highly uniform NPL waveguides, we demonstrate the first account of optical gain in only few NPL layers where the film thickness is precisely controlled. Our results show that NPLs as well as other types of NCs can benefit from this sequential liquid-air interface deposition technique for creation of two- or three-dimensional (3D) pure or hybrid NC superstructures. In the case of anisotropic NCs such as NPLs, the particle orientation can also be used in principle as an additional degree of freedom to choose and control in each layer during their 3D construction.

## 2. Results and Discussion

### 2.1. Multilayered Self-Assembly of NPLs

In this work, CdSe NPLs of 4.5 monolayer (ML) atomic thickness (having 5 Cd and 4 Se crystal planes alternatingly) were synthesized according to the previous literature.[6,7] Also, CdSe/Cd$_{0.25}$Zn$_{0.75}$S core/alloyed-shell NPLs were synthesized *via* hot-injection technique, as previously reported by our group (see Experimental Section),[9,26] and were dispersed in hexane. **Figure 1**a displays their transmission electron microscopy (TEM) images. The NPLs have square-like lateral shape with a side length of 17.8 ± 1.5 nm and a thickness of ~4.1 nm as determined from TEM imaging (see Figure S1). The absorbance and photoluminescence (PL) spectra of these NPLs are presented in Figure 1b. The first and second excitonic peaks reside at 631 and 573 nm, respectively, whereas the PL peak is at 642 nm.

To deposit these NPLs into thickness-controlled thin films, we modified our original self-assembly procedure, which we previously reported,[21] and applied it with the modifications sequentially as many times as needed to reach a certain targeted number of NPL layers in the construction. The modified technique is schematically illustrated in **Figure 2**a: Starting with precleaned blank substrates treated with 1H,1H,2H,2H-perfluorodecyltriethoxysilane (see Experimental Section) or continuing with previously deposited substrates, we fully immerse the substrates into diethylene glycol (DEG), which functions as the subphase. The hexane solution of NPLs is then dropped onto DEG, resulting in the liquid interface being almost fully covered with a membrane of NPLs after hexane is fully evaporated. To create and maintain a compact membrane, we adopted using silicone oil as a suspension material to create surface pressure, which helps compress the deposited nanoparticles, yielding a close-packed NPL film, as was previously used for preparing a close-packed film of NCs.[27] In our case, we dropped carefully the silicone oil dissolved in hexane from one edge to compress the membrane (see Experimental Section). Afterwards, the subphase is slowly drained with the help of a peristaltic pump at a rate of 50 μL/min, corresponding to ~260 nm descent of liquid level per s in the teflon dish of 64 mm in diameter used for the deposition. As the subphase is taken out, the membrane of NPLs eventually sinks onto the substrates while DEG is drawn away from the substrates. The surface treatment, which renders the substrates hydrophobic, as well as the tilted placement of the substrates help facilitating the repulsion of the subphase during the transfer of the NPL membrane onto the substrate. After the draining is complete, the residual DEG, if any, on the substrates is evaporated under vacuum at room temperature.

At the end of this procedure, one full monolayer of NPLs is deposited onto multiple substrates, as seen in the scanning electron microscopy (SEM) image of one of the substrates shown in Figure 2b (also see Figure S2c). The NPLs form a compact monolayer with excellent uniformity, with no visible multilayers or aggregation. Furthermore, owing to the compression by silicone oil, the deposited NPLs are densely packed, yielding near-unity surface coverage. Without the assistance of silicone oil, only a fraction of the NPLs on the membrane sticks to the surface during the transfer process, resulting in sub-monolayer coverage (see Figure S2). To further demonstrate the applicability of this technique to cover arbitrarily large areas, we deposited a single monolayer of NPLs onto a 4-inch fused silica wafer. The photograph in Figure 2c displays the luminescence across the wafer under UV excitation, where an area of about 80 $cm^2$ is uniformly deposited with a monolayer of NPLs.

In order to create thicker films with multiple NPL layers, this self-assembly procedure is repeated on the same previously deposited substrate as many times as desired. The resulting films are robust and maintain their strong emission and high uniformity. A cross-sectional TEM image taken from an exemplary case of 11 NPL layers, seen in Figure 2d, reveals that the NPL layers remain intact and do not introduce roughness as they build up. As can be clearly seen in the image, all of the NPL layers are distinctly visible, separated by their surface ligands. The uniformity of these layers continues to persist even over large areas (see Figure S3). Atomic force microscopy (AFM) imaging shows that the roughness of these multilayer films is limited to at most a few nm's even for the thickest NPL films (Figure S4b).

In addition, ellipsometric measurements on the thicknesses of multilayered NPL films having different numbers of layers verify the linear trend between the film thickness and the number of NPL layers as seen in Figure S4a. We extracted from the slope of the linear fit that each deposited monolayer adds up 7.0 nm to the film thickness. Since the thickness of the core/shell NPL structure is ~4.1 nm, the ligand brush between the NPL layers is deduced to be ~2.9 nm thick, which indicates a minor interdigitation between the ligand brushes of oleic acid/oleylamine. This interdigitation would allow the NPLs to stick together via van der Waals interaction, without having to deposit an oppositely charged linker molecule between the monolayers, as is commonly employed for layer-by-layer NC deposition.[28–31] Not having to rely on the deposition of oppositely charged particles adds up to the versatility of our multilayer deposition technique as the thickness of such oppositely charged bilayers can be altered by the ambient conditions,[32] or by unintentional diffusion of charged particles into the previously deposited inner layers during dipping steps, which leads to superlinear thickness growth.[33] Our self-assembly technique thus presents an alternative, robust approach to the existing methods of NC self-assembly at liquid interface,[34–37] while maintaining large-area uniformity and precision in film thickness.

**2.2. Optical Gain Measurements**

To characterize the NPL planar slabs in terms of their optical gain performance, we deposited NPL multilayers through the aforementioned construction procedure onto fused silica substrates with a varying number ($n$) of NPL layers up to $n = 15$. We were able to observe amplified spontaneous emission (ASE) from these films with $n = 6$ and above, whereas no ASE has been observed for $n \leq 5$. We therefore present our results for the films ranging from $n = 6$ (42 nm thick) to $n = 15$ (105 nm thick). We use a stripe excitation configuration for the ASE measurements, where the optical pump is normally incident on the specimen and the PL emission is collected from the side (see Experimental Section).[38] Our incident pump has 400 nm wavelength with 1 kHz pulse rate and ~110 fs pulse width. The PL emission of the NPL planar waveguide is collected from the side of the substrate with the help of an optical fiber while the pump fluence is varied.

The PL spectra were collected at various pump fluences for each $n$. These spectra are shown in **Figure 3**a for $n = 6$, 9 and 15. ASE is evident in all three cases from the emergence of the second emission feature beyond a threshold intensity, as well as the superlinear increase of the integrated emission intensity (Figure 3b). The per-pulse gain thresholds for these three cases have been determined to be 31.3, 12.8 and 7.5 $\mu J\ cm^{-2}$ for $n = 6$, 9 and 15, respectively. The tendency of the gain threshold to drop with increasing $n$ is also seen in Figure 3c, where the thresholds for the films of all the thicknesses from $n = 6$ to 15 are plotted. The gradual drop of the gain threshold continues up to $n = 15$, for which it is 7.5 $\mu J\ cm^{-2}$. This value is also on par with the reported gain thresholds of CdSe-based core/shell NPLs synthesized with hot-injection technique.[9,39–41]

Another notable observation is the evolution of the spectral position of the ASE peak with respect to the spontaneous emission for varied $n$. In Figure 3d is plotted one PL spectrum for each $n$ at pump intensities that are slightly above the respective gain thresholds. For the thinnest slabs the ASE peak displays little or no shift with respect to the spontaneous emission peak. As $n$ keeps increasing beyond 9, however, a gradual increase in the red shift of this peak is observed until $n = 15$ (see Figure 3d), where the ASE peak saturates at around 664 nm (see Figure 3d and Table S1) and its red shift is determined to be 18.2 nm. Similar change in the shift of ASE peak with film thickness has previously been observed with the thin films composed of other types of emitters, and was attributed to the change in the critical wavelength of propagation in the planar waveguides as the film thickness is varied.[17,18] In our case, this shift of the ASE peak allows for fine tuning of the optical gain wavelength across a 17 nm wide spectral band (see Table S1) simply by precisely adjusting the NPL waveguide thickness, which is controlled by the sequential NPL deposition.

## 2.3. Mode Analysis on NPL Planar Waveguides

To elaborate on the observed characteristics of the NPL slabs, we employed a numerical analysis of guided modes within the NPL layer. Approximating the NPL multilayer as a homogeneous medium, our structure can be modeled as an asymmetric planar waveguide confined between semi-infinite fused silica and air (**Figure 4**a). As the refractive index of the NPL layer, which is measured as $n_2 = 1.96$ via ellipsometry, is greater than those of air ($n_1 = 1$) and fused silica ($n_3 = 1.45$), this configuration will allow propagation of the supported modes through the NPL layer via total internal reflection. However, the existence of these modes requires a minimum thickness of the slab waveguide, which can be calculated using the well-established planar waveguide formalism. Accordingly, the minimum thickness $t_c$ (critical thickness) of the waveguide necessary for the $m^{th}$ order transverse electric mode (TE$_m$) is given by

$$t_c = \frac{\lambda}{2\pi\sqrt{n_2^2 - n_3^2}} \left[ m\pi + tan^{-1}\left(\left(\frac{n_3^2 - n_1^2}{n_2^2 - n_3^2}\right)^{1/2}\right) \right] \quad (1)$$

where $\lambda$ is the wavelength of the mode in free space.[42] For $\lambda = 650$ nm, **Equation 1** reveals that the minimum thickness required for the TE$_0$ mode is 52.7 nm, which is in close proximity to the experimental onset of ASE in our multilayered NPLs observed with 6 layers (thickness of 42 nm). The slight discrepancy between the experimental critical thickness and the calculated minimum thickness at which ASE is observed could be due to the effective refractive index of the NPL medium being modified under hard pumping conditions. The fact that optical gain is observed at 42 nm shows that the actual refractive index of the NPL film is greater than 1.96 and nonzero modal confinement is possible at this thickness. For numerical convenience, we therefore take $n_2 = 2.10$ for mode calculations that will follow. This refractive index is not far off from the one measured by ellipsometry (n$_2$ = 1.96), and numerically yields $t_c = 41.2$ nm.

The calculated mode intensity profile of $TE_0$ mode is also plotted in Figure 4a for 6-layered NPL waveguide. Due to the ultra-thin NPL film, only a small portion of the propagating field is confined within the actual gain medium. The field confinement factor, $\Gamma$, is calculated as $6.3\times10^{-3}$ for $n = 6$. This factor monotonously increases with additional NPL layers deposited, as seen in Figure 4b. More notably, it undergoes a significant jump of 4 orders of magnitude from $n = 5$ to $n = 6$, which is in accordance with the onset of optical gain. By increasing the thickness of the gain medium even further, the optical gain is facilitated by the stronger confinement of the optical mode into the NPL slab, hence the reduction in the gain threshold.

To account for the spectral shifting of the ASE peak with film thickness, we consider the variance of the mode confinement with mode wavelength. Previously, observing a similar trend in the ASE peak of optically pumped polystyrene films, Calzado *et al.* argued qualitatively that each vibronic mode experiences a different loss at different film thicknesses. As a result, the propagation losses for the modes having longer wavelengths are enhanced compared to those supported at shorter wavelengths, which explains the blue-shift of the ASE peak at thinner polystyrene films.[18] A similar argument can be made for our NPL constructs as well, since the mode confinement near the critical thickness is highly sensitive to the wavelength, as seen in Figure 4c. Furthermore, NPLs as well as other classes of NCs provide a certain gain bandwidth, which determines the spectral range of gain that can be observed from the material. In the generic case, this gain band is not flat, i.e. the material gain coefficient $G$ is a function of $\lambda$. The total unsaturated gain coefficient will then be given by $g_0(\lambda) = G_0(\lambda)\Gamma(\lambda)$, neglecting the losses. When the gain is saturated, only the modes around the maximum of $g_0(\lambda)$ will survive and contribute to the gain. Therefore, the ASE peak at the gain saturation can be estimated by calculating the maximum of $G_0(\lambda)\Gamma(\lambda)$.

To test if this simple formalism can explain the observed ASE shift, we model the unsaturated material gain as a generalized Gaussian function:

$$G_0(\lambda) = G_{0c} e^{-\left(\frac{|\lambda - \lambda_c|}{\alpha}\right)^{\beta}} \quad (2)$$

where $G_{0c}$ is the unsaturated gain coefficient at the peak wavelength $\lambda_c$, and $\beta$ and $\alpha$ are parameters related to the shape and width of the distribution, respectively. $\lambda_c$ is fixed at 665 nm since it is the peak of the most red-shifted ASE feature among all the tested multilayered films. The parameters $\beta$ and $\alpha$ are swept to find the parameter couple that minimizes the difference between the experimentally determined ASE peaks and the calculated ones. The optimum parameters found this way are $\beta = 2.0$ and $\alpha = 64$, for which **Equation 2** is plotted in Figure 4c. The calculated ASE peaks are plotted in Figure 4d, along with the experimentally determined peaks. We see that, apart from the ASE peak for $n = 6$, the assumed $G_0(\lambda)$ can estimate the trend in the ASE peak shift fairly well. The discrepancy at $n = 6$ is most likely due to the actual $G_0(\lambda)$ having a finite range that does not include 630 nm. Therefore, the blue shift of the ASE peak at thinner films is limited by the finite gain band of the material. It is also worth noting that, because for very thick films $\Gamma(\lambda) \approx 1$ throughout the gain band, this model can readily predict that with increasing thickness, the ASE peak will converge to $\lambda_c$. Indeed, the experimental ASE peaks seem to be saturating towards a definite red shift around $\lambda = 665$ nm. Although this type of red shift is expected for NCs having type-I band alignment[43], here we show that it is possible to obtain non-shifting ASE from type-I NCs by using their ultra-thin films.

## 3. Conclusion

In summary, we have created planar waveguides made out of the superstructures of our NPLs laid in face-down orientation by a sequentially-repeated liquid interface self-assembly technique, which is applicable to large scales. The multilayered NPL films prepared with this technique have close packing and excellent uniformity over many tens of cm$^2$ large areas. These NPL constructs can work as optical slab waveguides, of which optical confinement is revealed as a function of the number of NPL layers, or equivalently, their film thickness, at a monolayer precision. Our findings show that these self-assembled NPL slabs can also be used as optically active planar waveguides with precise thickness-controlled gain, which can find applications in on-chip active photonic devices requiring in-plane waveguided light. The gradually decreasing gain threshold observed with increasing number of layers is revealed to result from the optical confinement increasing with the deposited NPL layers.

In the case of electrically-driven NC films, the organic surfactants in NC films act as insulating barriers that impede charge or heat transfer, limiting their use in electroluminescent devices. The ultimate thickness of our NPL film inducing optical gain is as small as ~40 nm, with only 6 layers of NPLs, may result in significant reduction of organic barriers that should be overcome by the electric current in the vertical direction. This work may therefore serve as a foundation for developing the optically active media of electrically-driven NPL lasers without having to switch to inorganic ligands. The technique presented in this work can be readily extended to other types of NCs to create large-area hybrid NC superstructures, which can be useful for device fabrication. In addition, the anisotropy of the non-spherical NCs can also be used as a degree of freedom to create NC multilayers having different orientations in different layers. Therefore, this multilayered self-assembly technique paves the way for exploration of a rich variety of large-area bottom-up NC superstructure construction.

## 4. Experimental Section

*Synthesis of CdSe/Cd$_{0.25}$Zn$_{0.75}$S core/shell NPLs.* CdSe/CdZnS core/alloyed-shell NPLs have been synthesized with some modifications by using the hot-injection shell synthesis recipe reported in our recent works.[9,26] Cd-acetate (17.25 mg), Zn-acetate (41.25 mg), 1-octadecene (ODE, 7 mL), oleic acid (0.70 mL) and 4.5 ML CdSe core in hexane (1 mL, having absorbance of 1.5 at 350 nm along 1 cm-long optical path) have been loaded in quartz flask (50 mL) and stirred for 90 min at room temperature, for 30 min at 85 °C under vacuum environment. Then, oleylamine (0.7 mL) has been injected into the flask at 90 °C under argon gas and the temperature of the solution has been increased to 300 °C. The injection of ODE-octanethiol mixture that is prepared in glovebox using ODE (4 mL) and octanethiol (105 µL) have been started at 165 °C using 10 mL h$^{-1}$ rate of syringe pump until the temperature reached 280 °C, and then the rate has been decreased to 4 mL h$^{-1}$. Before cooling to room temperature with cold water, the solution is kept for an hour at 300°C. Hexane (5 mL) is added at room temperature and the solution is cleaned four times with absolute ethanol. Finally, the precipitated NPLs is dissolved in hexane.

*Multi-layered self-assembly of CdSe/Cd$_{0.25}$Zn$_{0.75}$S*: Precleaned fused silica substrates having 13×13 mm$^2$ size or silicon substrates with 10×10 mm$^2$ size are treated with the vapor of 1H,1H,2H,2H-perfluorodecyltriethoxysilane at 200 °C under nitrogen environment for 30 minutes in order to render their surfaces hydrophobic.[44] Treated substrates are rinsed first in hexane, then acetone and isopropanol, and next attached to a stage tilted by ~10° and finally immersed into diethylene glycol (DEG) (Merck) contained in a teflon dish. One end of the tubing of a peristaltic pump is inserted into DEG. NPL solution of hexane is dropped onto the DEG interface, which quickly spreads across the surface and evaporates in a few seconds. After the NPL solution is fully evaporated, one drop of silicone oil (silicone elastomer, Sylgard 184) dissolved in hexane (2 mg/mL) is dropped from close to the edge of the DEG interface, which compresses the NPL membrane. As the peristaltic pump slowly drains the DEG, the NPL membrane on the interface eventually settles onto the substrates. Any residual DEG on the substrates is dried under vacuum at room temperature. For multilayered deposition, the same substrates are submerged back into DEG for another deposition cycle. The number of depositions carried out on a substrate determines the number of NPL layers on it.

*Optical gain measurements*: Multilayered NPLs deposited on fused silica have been placed in front of a pulsed laser beam at normal incidence, whose intensity is adjusted with a neutral density filter on the optical path. A Ti:Sapphire laser amplifier (Spitfire Pro) generates laser pulses at 800 nm with a pulse width of ~110 fs and a repetition rate of 1 kHz. The wavelength of these pulses is converted to 400 nm with a frequency-doubling barium borate crystal. The residual 800 nm light is filtered out with a short pass filter. The 400 nm beam is focused along one dimension onto the sample with a cylindrical lens to obtain a stripe excitation having a width of ~150 μm. The emission of the sample is collected from the side with an optical fiber. The spectra are recorded with an optical spectrometer (Maya 2000, Ocean Optics).

*Mode profile calculations*: The modal analyses were carried out using a commercially available mode solution software package (Lumerical FDTD) with two-dimensional layout. The refractive indices of CQW and fused silica used in the simulations were 2.10 and 1.45, respectively. The confinement factor was calculated based on the ratio of the fundamental transverse electric mode energy confined within the active NPL film to the total mode energy.

**Supporting Information**
Supporting Information is available from the Wiley Online Library or from the author.

**Acknowledgements**
The authors acknowledge the financial support from the Singapore National Research Foundation under the program NRF-NRFI2016-08 and in part from TUBITAK 115E679. The authors thank Mr. Mustafa Guler for TEM imaging of the as-synthesized NPLs and preparation of the TEM cross-sectional sample, Mr. Semih Bozkurt for his support on the AFM characterization, Dr. Gokce Celik for her help on the ellipsometric measurements, Mr. Emre Unal for his assistance in photography of the large-area sample, and Dr. Kivanc Gungor for fruitful discussions. O.E. acknowledges TUBITAK for the financial support through BIDEB 2211 program. H.V.D. gratefully acknowledges support from TÜBA.

**References**

[1]    V. I. Klimov, A. A. Mikhailovsky, S. Xu, A. Malko, J. A. Hollingsworth, C. A. Leatherdale, H. J. Eisler, M. G. Bawendi, *Science* **2000**, *290*, 314.

[2]    H. Htoon, J. A. Hollingworth, A. V. Malko, R. Dickerson, V. I. Klimov, *Appl. Phys. Lett.* **2003**, *82*, 4776.

[3]    C. Dang, J. Lee, C. Breen, J. S. Steckel, S. Coe-Sullivan, A. Nurmikko, *Nat. Nanotechnol.* **2012**, *7*, 335.

[4]    C. She, I. Fedin, D. S. Dolzhnikov, A. Demortière, R. D. Schaller, M. Pelton, D. V. Talapin, *Nano Lett.* **2014**, *14*, 2772.

[5]    B. Guzelturk, Y. Kelestemur, K. Gungor, A. Yeltik, M. Z. Akgul, Y. Wang, R. Chen, C.

Dang, H. Sun, H. V. Demir, *Adv. Mater.* **2015**, *27*, 2741.

[6]   S. Ithurria, M. D. Tessier, B. Mahler, R. P. S. M. Lobo, B. Dubertret, A. L. Efros, *Nat. Mater.* **2011**, *10*, 936.

[7]   M. D. Tessier, C. Javaux, I. Maksimovic, V. Loriette, B. Dubertret, *ACS Nano* **2012**, *6*, 6751.

[8]   B. T. Diroll, D. V. Talapin, R. D. Schaller, *ACS Photonics* **2017**, *4*, 576.

[9]   Y. Altintas, K. Gungor, Y. Gao, M. Sak, U. Quliyeva, G. Bappi, E. Mutlugun, E. H. Sargent, H. V. Demir, *ACS Nano* **2019**, *13*, 10662.

[10]  B. Guzelturk, M. Pelton, M. Olutas, H. V. Demir, *Nano Lett.* **2019**, *19*, 277.

[11]  B. Guzelturk, Y. Kelestemur, M. Olutas, S. Delikanli, H. V. Demir, *ACS Nano* **2014**, *8*, 6599.

[12]  C. She, I. Fedin, D. S. Dolzhnikov, P. D. Dahlberg, G. S. Engel, R. D. Schaller, D. V. Talapin, *ACS Nano* **2015**, *9*, 9475.

[13]  Q. Li, Q. Liu, R. D. Schaller, T. Lian, *J. Phys. Chem. Lett.* **2019**, *10*, 1624.

[14]  M. Sak, N. Taghipour, S. Delikanli, S. Shendre, I. Tanriover, S. Foroutan, Y. Gao, J. Yu, Z. Yanyan, S. Yoo, C. Dang, H. V. Demir, *Adv. Funct. Mater.* **2020**, *30*, 1907417.

[15]  G. J. Denton, N. Tessler, M. A. Stevens, R. H. Friend, *Adv. Mater.* **1997**, *9*, 547.

[16]  X. Long, M. Grell, A. Malinowski, D. D. C. Bradley, M. Inbasekaran, E. P. Woo, *Opt. Mater. (Amst).* **1998**, *9*, 70.

[17]  A. K. Sheridan, G. A. Turnbull, A. N. Safonov, I. D. W. Samuel, *Phys. Rev. B - Condens. Matter Mater. Phys.* **2000**, *62*, 929.

[18]  E. M. Calzado, J. M. Villalvilla, P. G. Boj, J. A. Quintana, M. A. Díaz-García, *J. Appl. Phys.* **2005**, *97*, 093103.

[19]  B. T. Diroll, N. J. Greybush, C. R. Kagan, C. B. Murray, *Chem. Mater.* **2015**, *27*, 2998.

[20]  F. Feng, L. T. Nguyen, M. Nasilowski, B. Nadal, B. Dubertret, A. Maître, L. Coolen, *ACS Photonics* **2018**, *5*, 1994.

[21] O. Erdem, K. Gungor, B. Guzelturk, I. Tanriover, M. Sak, M. Olutas, D. Dede, Y. Kelestemur, H. V. Demir, *Nano Lett.* **2019**, *19*, 4297.

[22] T. Franzl, T. A. Klar, S. Schietinger, A. L. Rogach, J. Feldmann, *Nano Lett.* **2004**, *4*, 1599.

[23] T. Ozel, S. Nizamoglu, M. A. Sefunc, O. Samarskaya, I. O. Ozel, E. Mutlugun, V. Lesnyak, N. Gaponik, A. Eychmuller, S. V. Gaponenko, H. V. Demir, *ACS Nano* **2011**, *5*, 1328.

[24] Y. Chan, J. M. Caruge, P. T. Snee, M. G. Bawendi, *Appl. Phys. Lett.* **2004**, *85*, 2460.

[25] J. Roh, Y. S. Park, J. Lim, V. I. Klimov, *Nat. Commun.* **2020**, *11*, 271.

[26] B. Liu, Y. Altintas, L. Wang, S. Shendre, M. Sharma, H. Sun, E. Mutlugun, H. V. Demir, *Adv. Mater.* **2020**, *32*, 1.

[27] H. Min, J. Zhou, X. Bai, L. Li, K. Zhang, T. Wang, X. Zhang, Y. Li, Y. Jiao, X. Qi, Y. Fu, *Langmuir* **2017**, *33*, 6732.

[28] G. Decher, *Science* **1997**, *277*, 1232.

[29] J. Roither, S. Pichler, M. V. Kovalenko, W. Heiss, P. Feychuk, O. Panchuk, J. Allam, B. N. Murdin, *Appl. Phys. Lett.* **2006**, *89*, 1.

[30] T. Ozel, P. L. Hernandez-Martinez, E. Mutlugun, O. Akin, S. Nizamoglu, I. O. Ozel, Q. Zhang, Q. Xiong, H. V. Demir, *Nano Lett.* **2013**, *13*, 3065.

[31] I. Suarez, R. Munoz, V. Chirvony, J. P. Martinez-Pastor, M. Artemyev, A. Prudnikau, A. Antanovich, A. Mikhailov, *IEEE J. Sel. Top. Quantum Electron.* **2017**, *23*, 1.

[32] T. J. Halthur, P. M. Claesson, U. M. Elofsson, *J. Am. Chem. Soc.* **2004**, *126*, 17009.

[33] P. T. Hammond, *AIChE J.* **2011**, *57*, 2928.

[34] A. Dong, J. Chen, P. M. Vora, J. M. Kikkawa, C. B. Murray, *Nature* **2010**, *466*, 474.

[35] K. Lambert, R. K. Čapek, M. I. Bodnarchuk, M. V. Kovalenko, D. Van Thourhout, W. Heiss, Z. Hens, *Langmuir* **2010**, *26*, 7732.

[36] N. Vogel, L. De Viguerie, U. Jonas, C. K. Weiss, K. Landfester, *Adv. Funct. Mater.*

**2011**, *21*, 3064.

[37] T. Paik, D. K. Ko, T. R. Gordon, V. Doan-Nguyen, C. B. Murray, *ACS Nano* **2011**, *5*, 8322.

[38] K. L. Shaklee, R. E. Nahory, R. F. Leheny, *J. Lumin.* **1973**, *7*, 284.

[39] Y. Altintas, U. Quliyeva, K. Gungor, O. Erdem, Y. Kelestemur, E. Mutlugun, M. V. Kovalenko, H. V. Demir, *Small* **2019**, *15*, 1.

[40] A. A. Rossinelli, H. Rojo, A. S. Mule, M. Aellen, A. Cocina, E. De Leo, R. Schäublin, D. J. Norris, *Chem. Mater.* **2019**, *31*, 9567.

[41] L. Zhang, H. Yang, B. Yu, Y. Tang, C. Zhang, X. Wang, M. Xiao, Y. Cui, J. Zhang, *Adv. Opt. Mater.* **2020**, *8*, 1.

[42] A. Yariv, P. Yeh, *Photonics: Optical Electronics in Modern Communications*, Oxford University Press, New York, NY, USA, **2007**.

[43] V. I. Klimov, S. A. Ivanov, J. Nanda, M. Achermann, I. Bezel, J. A. McGuire, A. Piryatinski, *Nature* **2007**, *447*, 441.

[44] M. Beck, M. Graczyk, I. Maximov, E. L. Sarwe, T. G. I. Ling, M. Keil, L. Montelius, *Microelectron. Eng.* **2002**, *61–62*, 441.

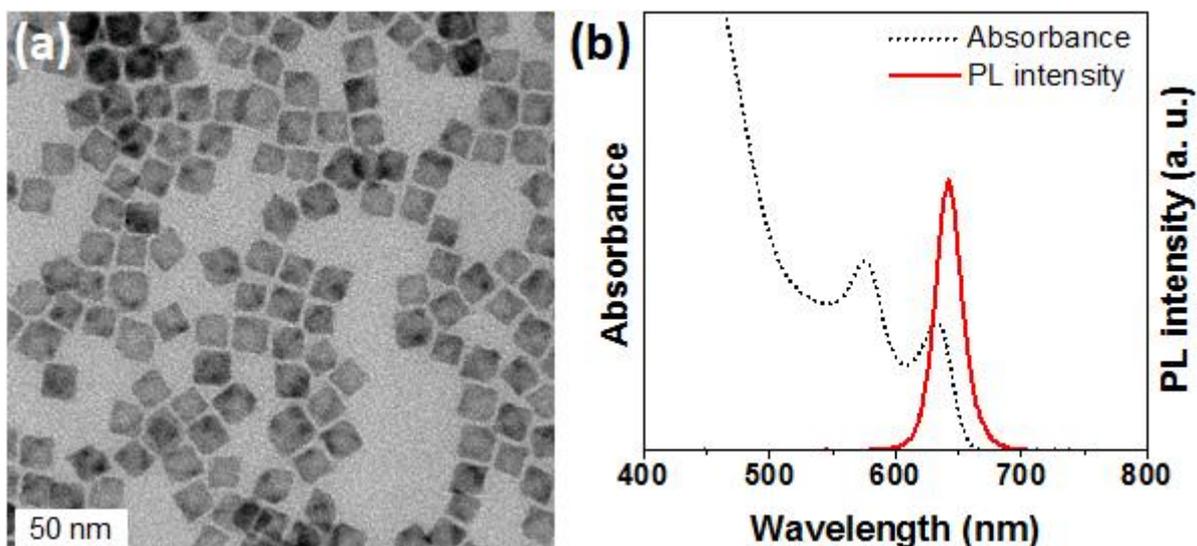

**Figure 1.** (a) Transmission electron micrographs of square-shaped CdSe/Cd$_{0.25}$Zn$_{0.75}$S core/shell NPLs. (b) Absorbance and photoluminescence spectra for the CdSe/Cd$_{0.25}$Zn$_{0.75}$S NPLs.

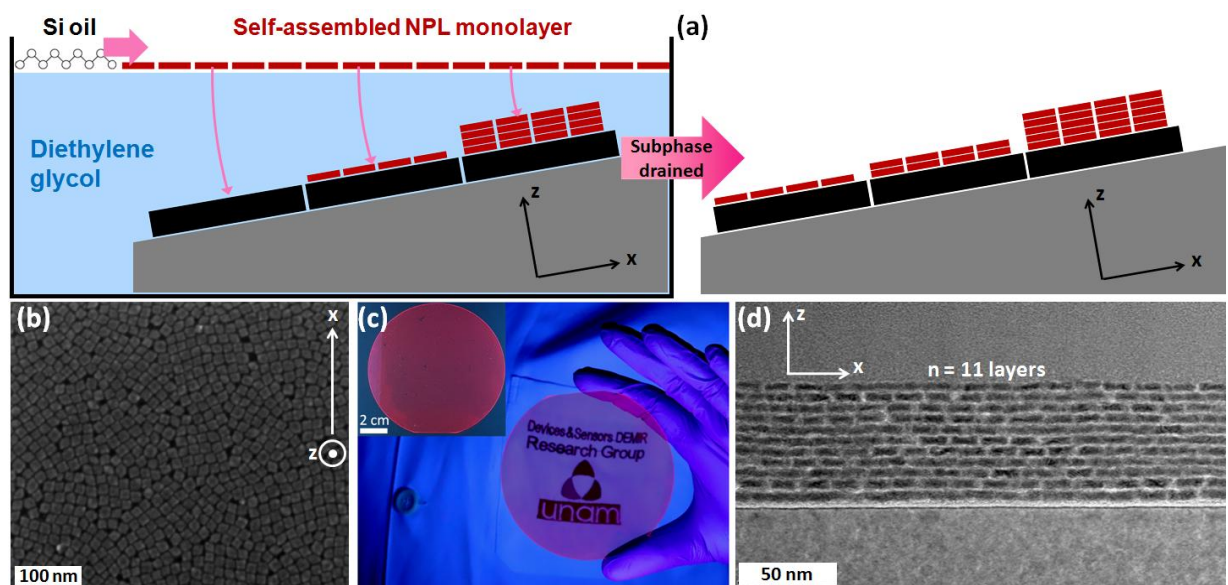

**Figure 2.** (a) Schematic demonstration of liquid-air interface self-assembly of the NPLs. Blank or predeposited substrates are inserted into a subphase of diethylene glycol. The NPL solution is dropped onto the subphase and quickly spreads across the liquid-air interface. Dropped silicone oil compresses the NPL membrane. After the subphase is drained, all the substrates are deposited with one additional layer of NPLs. (b) Scanning electron micrograph of one monolayer of NPLs deposited through liquid-air interface self-assembly. (c) Photograph of one monolayer of NPLs deposited onto 4-inch fused silica illuminated under UV light. Scale bar in the inset is 2 cm. (d) Cross-sectional TEM image of the 11 NPL monolayers sequentially deposited onto silicon. All the NPL layers are distinctly visible, separated by their surface ligands.

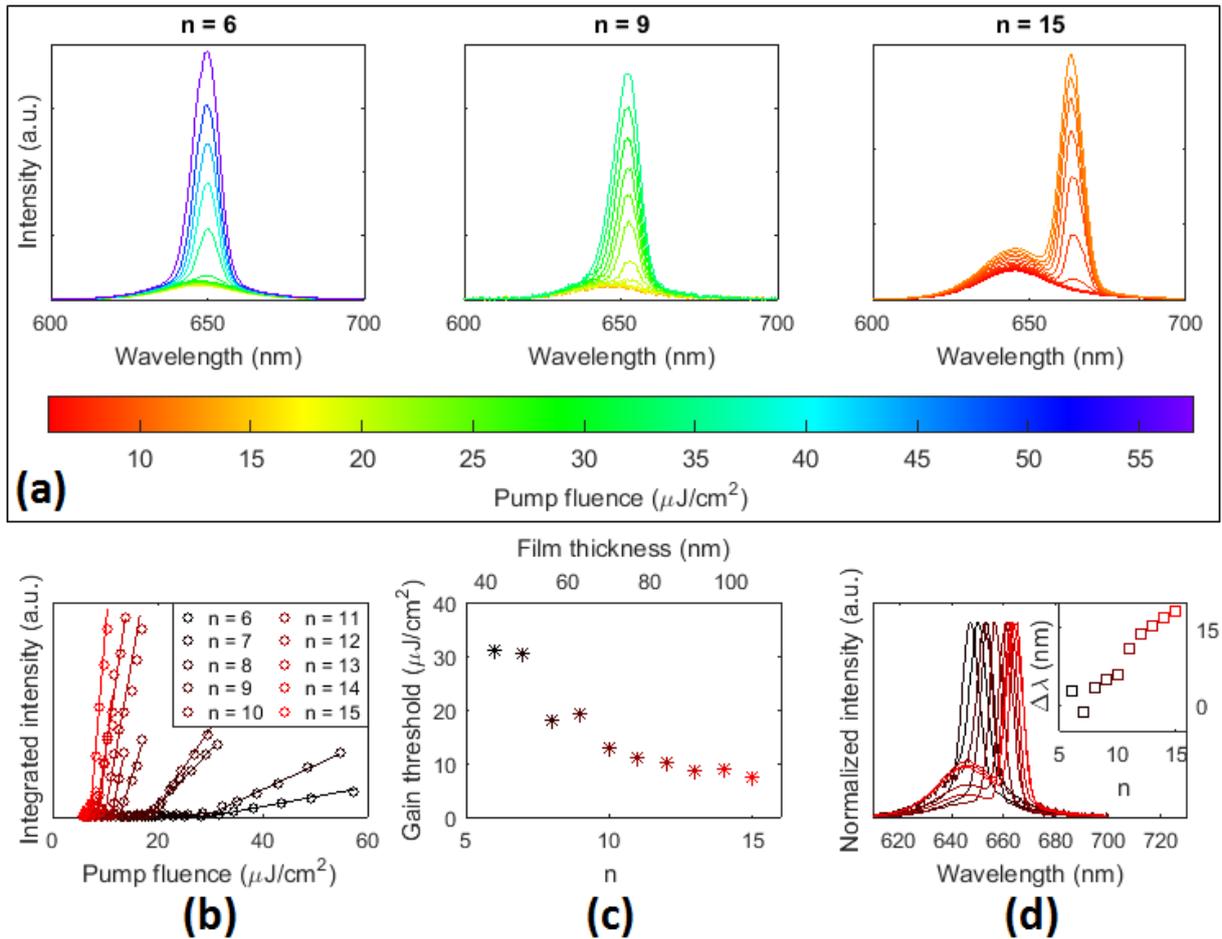

**Figure 3.** (a) Emission spectra of the NPL films having different number $n$ of layers excited with a pulsed laser at 400 nm: left, $n = 6$, center, $n = 9$ and right, $n = 15$. Colorbar at the bottom is common for all three plots. (b) Integrated intensity as a function of the pump fluence for all the NPL films from $n = 6$ to $n = 15$. (c) Evolution of the optical gain threshold with the number of layers $n$ as deduced from the data in panel b. (d) Shifting of the ASE peak with respect to the spontaneous emission for different $n$. Inset shows the difference between ASE and spontaneous emission peaks. The color coding is identical in panels b-d.

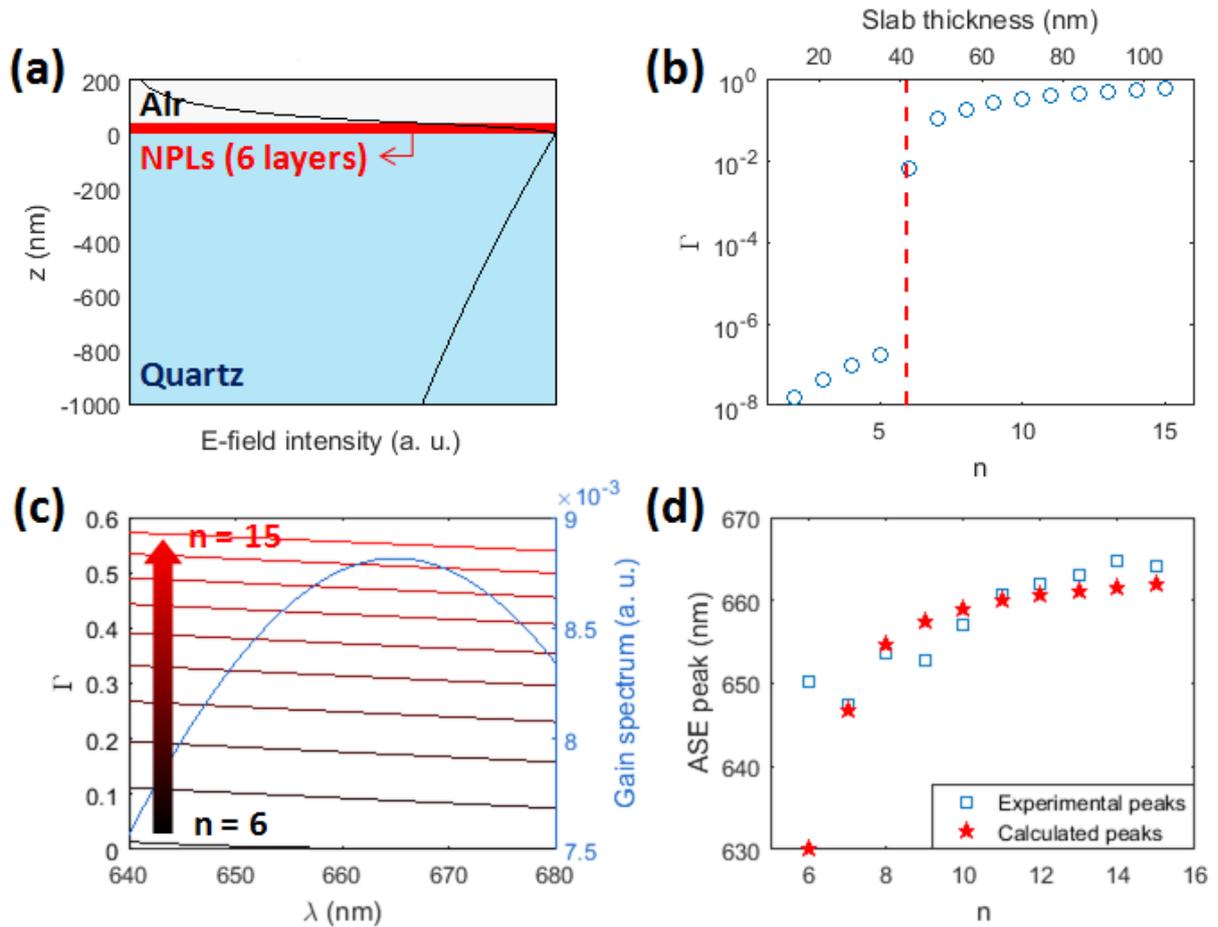

**Figure 4.** (a) Electric field profile of the fundamental TE mode numerically calculated for the NPL slab with $n = 6$. (b) Optical confinement factor $\Gamma$ calculated at 650 nm for different number of NPL layers (and slab thickness). Dashed line indicates the critical thickness of 41.2 nm for the existence of propagating modes within the slab. (c) Left axis: Variation of the confinement factor $\Gamma$ with wavelength for different numbers of layers. Right axis: Unsaturated gain spectrum $g(\lambda)$ for our NPLs estimated using the measured ASE peaks at different film thicknesses, modeled as a Gaussian centered at 665 nm with a FWHM of 106.6 nm (blue) (d) ASE peaks calculated as the maximum of $\Gamma(\lambda) \times g(\lambda)$ for each n (red stars), together with the experimentally measured ASE peaks (blue squares).

Supporting Information

**Orientation-Controlled Construction of Self-Assembled Superstructures of Atomically-Flat Nanocrystals: Pushing the Limits of Ultra-Thin Colloidal Gain Media**

*Onur Erdem, Sina Foroutan, Negar Gheslaghi, Burak Guzelturk, Yemliha Altintas and Hilmi Volkan Demir\**

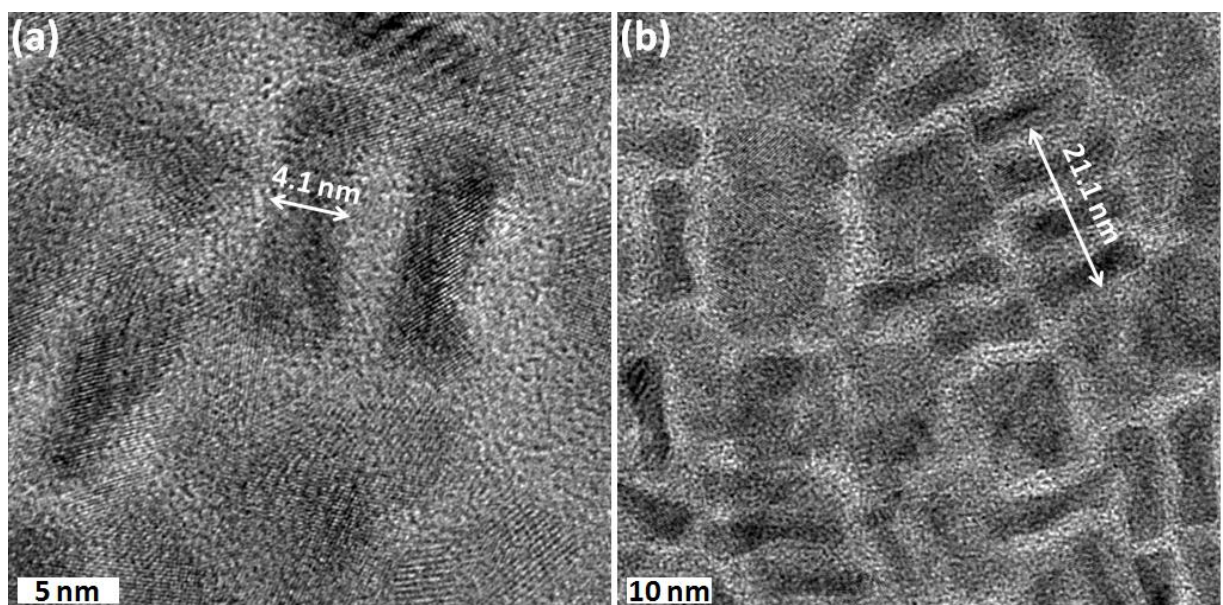

**Figure S1.** High-resolution transmission electron micrographs of CdSe/Cd$_{0.25}$Zn$_{0.75}$S core/alloyed-shell NPLs that are vertically oriented. (a) The thickness of a single NPL is measured as ~4.1 nm. (b) Measurements on chains of vertically oriented NPLs reveal a center-to-center distance of ~7.0 nm, in accordance with the ellipsometric measurements (see Figure S4a)

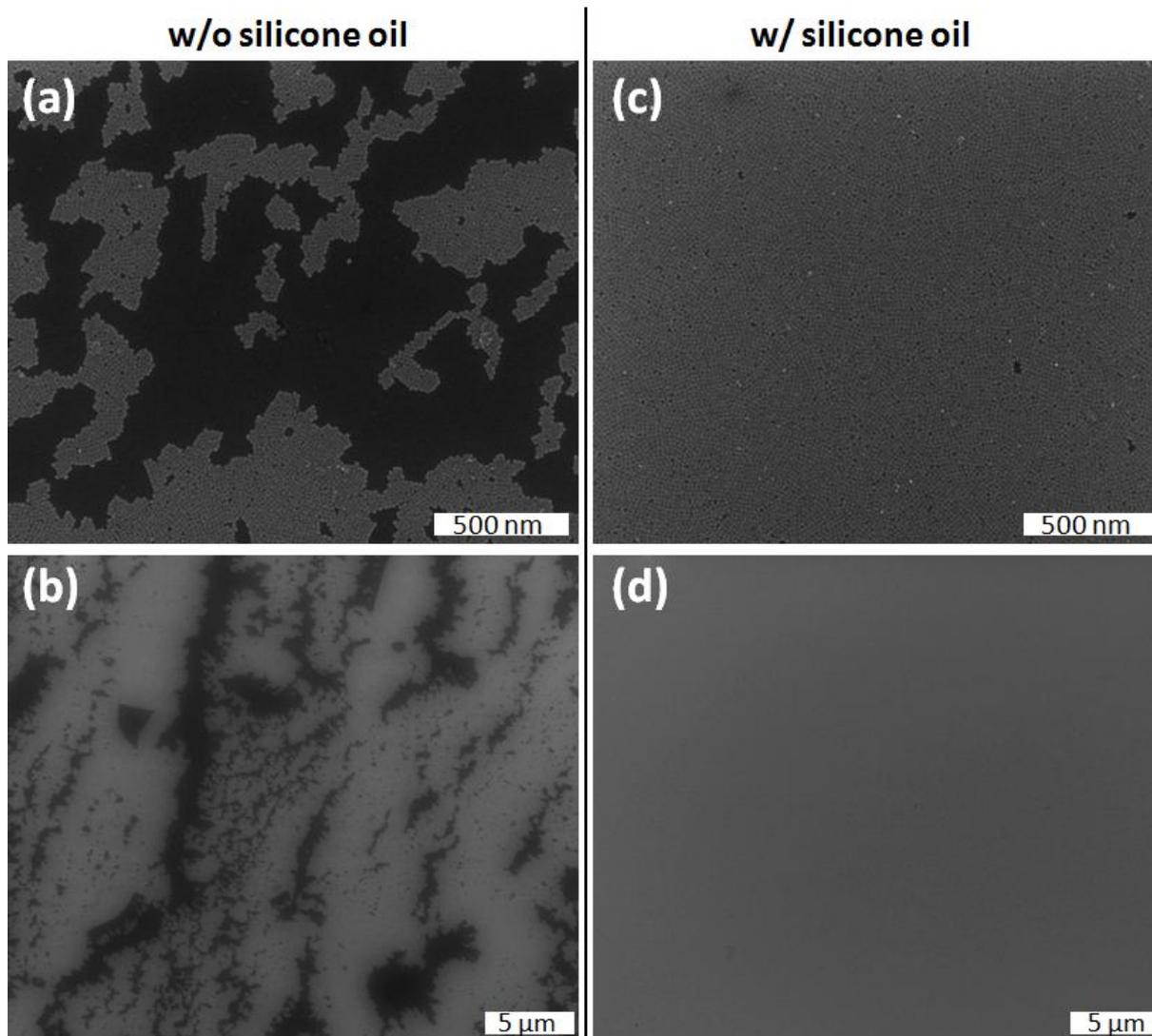

**Figure S2.** Scanning electron micrographs of self-assembled NPLs (a, b) without and (c, d) with silicone oil. In all panels, dark areas are void.

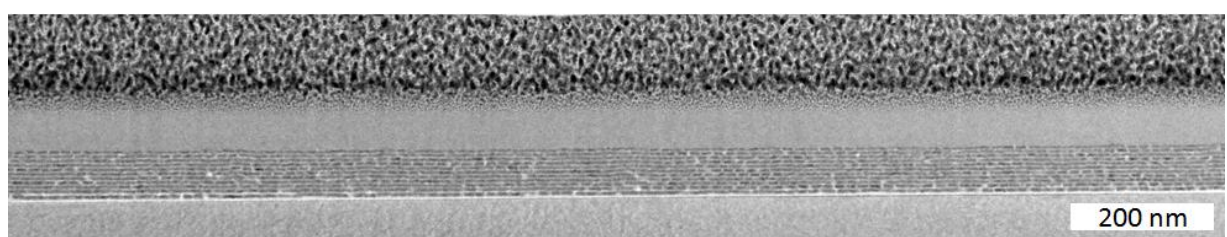

**Figure S3.** Cross-sectional transmission electron micrograph of 11-layered NPLs at lower magnification.

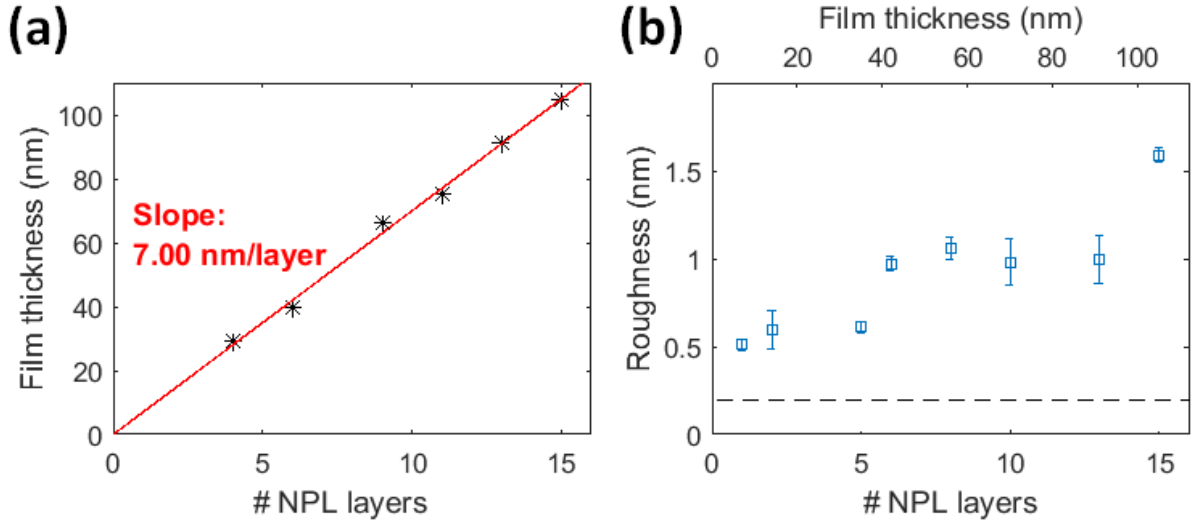

**Figure S4.** (a) Measurement of film thickness for the multilayered NPL constructs having different numbers of layers. The linear fit confirms 7.0 nm of thickness per NPL layer. (b) Surface roughness measurements of multilayer NPL films taken with atomic force microscopy over 5 different regions having an area of 2×2 μm$^2$. Dashed line shows the roughness of the bare fused silica substrate (~0.2 nm).

**Table S1.** Gaussian fitting parameters for the ASE spectra presented in Figure 2d. Peak and FWHM values for spontaneous emission (SE) and ASE features.

| # NPL layers | SE peak (nm) | SE width (nm) | ASE peak (nm) | ASE width (nm) |
|---|---|---|---|---|
| 6 | 647.6 | 31.8 | 650.2 | 7.2 |
| 7 | 648.8 | 32.6 | 647.4 | 7.4 |
| 8 | 650.3 | 29.1 | 653.7 | 5.9 |
| 9 | 647.8 | 30.3 | 652.7 | 7.6 |
| 10 | 651.3 | 35.0 | 657.1 | 6.2 |
| 11 | 649.7 | 35.5 | 660.8 | 5.9 |
| 12 | 648.0 | 34.0 | 661.9 | 5.7 |
| 13 | 647.7 | 32.3 | 663.0 | 5.5 |
| 14 | 647.7 | 33.0 | 664.8 | 5.5 |
| 15 | 645.9 | 31.1 | 664.1 | 7.2 |